\PassOptionsToPackage{unicode}{hyperref}
\PassOptionsToPackage{hyphens}{url}
\PassOptionsToPackage{dvipsnames,svgnames,x11names}{xcolor}
\documentclass[
  letterpaper,
  DIV=11,
  numbers=noendperiod]{scrartcl}
\usepackage{xcolor}
\usepackage{amsmath,amssymb}
\usepackage{iftex}
\ifPDFTeX
  \usepackage[T1]{fontenc}
  \usepackage[utf8]{inputenc}
  \usepackage{textcomp} % provide euro and other symbols
\else % if luatex or xetex
  \usepackage{unicode-math} % this also loads fontspec
  \defaultfontfeatures{Scale=MatchLowercase}
  \defaultfontfeatures[\rmfamily]{Ligatures=TeX,Scale=1}
\fi
\usepackage{lmodern}
\ifPDFTeX\else
\fi
\IfFileExists{upquote.sty}{\usepackage{upquote}}{}
\IfFileExists{microtype.sty}{% use microtype if available
  \usepackage[]{microtype}
  \UseMicrotypeSet[protrusion]{basicmath} % disable protrusion for tt fonts
}{}
\makeatletter
\@ifundefined{KOMAClassName}{% if non-KOMA class
  \IfFileExists{parskip.sty}{%
    \usepackage{parskip}
  }{% else
    \setlength{\parindent}{0pt}
    \setlength{\parskip}{6pt plus 2pt minus 1pt}}
}{% if KOMA class
  \KOMAoptions{parskip=half}}
\makeatother
\makeatletter
\ifx\paragraph\undefined\else
  \let\oldparagraph\paragraph
  \renewcommand{\paragraph}{
    \@ifstar
      \xxxParagraphStar
      \xxxParagraphNoStar
  }
  \newcommand{\xxxParagraphStar}[1]{\oldparagraph*{#1}\mbox{}}
  \newcommand{\xxxParagraphNoStar}[1]{\oldparagraph{#1}\mbox{}}
\fi
\ifx\subparagraph\undefined\else
  \let\oldsubparagraph\subparagraph
  \renewcommand{\subparagraph}{
    \@ifstar
      \xxxSubParagraphStar
      \xxxSubParagraphNoStar
  }
  \newcommand{\xxxSubParagraphStar}[1]{\oldsubparagraph*{#1}\mbox{}}
  \newcommand{\xxxSubParagraphNoStar}[1]{\oldsubparagraph{#1}\mbox{}}
\fi
\makeatother

\usepackage{longtable,booktabs,array}
\usepackage{calc} % for calculating minipage widths
\usepackage{etoolbox}
\makeatletter
\patchcmd\longtable{\par}{\if@noskipsec\mbox{}\fi\par}{}{}
\makeatother
\IfFileExists{footnotehyper.sty}{\usepackage{footnotehyper}}{\usepackage{footnote}}
\makesavenoteenv{longtable}
\usepackage{graphicx}
\makeatletter
\newsavebox\pandoc@box
\newcommand*\pandocbounded[1]{% scales image to fit in text height/width
  \sbox\pandoc@box{#1}%
  \Gscale@div\@tempa{\textheight}{\dimexpr\ht\pandoc@box+\dp\pandoc@box\relax}%
  \Gscale@div\@tempb{\linewidth}{\wd\pandoc@box}%
  \ifdim\@tempb\p@<\@tempa\p@\let\@tempa\@tempb\fi% select the smaller of both
  \ifdim\@tempa\p@<\p@\scalebox{\@tempa}{\usebox\pandoc@box}%
  \else\usebox{\pandoc@box}%
  \fi%
}
\def\fps@figure{htbp}
\makeatother

\NewDocumentCommand\citeproctext{}{}

\makeatletter
 \let\@cite@ofmt\@firstofone
 \def\@biblabel#1{}
 \def\@cite#1#2{{#1\if@tempswa , #2\fi}}
\makeatother
\newlength{\cslhangindent}
\newlength{\csllabelwidth}
\newenvironment{CSLReferences}[2] % #1 hanging-indent, #2 entry-spacing
 {\begin{list}{}{%
  \setlength{\itemindent}{0pt}
  \setlength{\leftmargin}{0pt}
  \setlength{\parsep}{0pt}
  \ifodd #1
   \setlength{\leftmargin}{\cslhangindent}
   \setlength{\itemindent}{-1\cslhangindent}
  \fi
  \setlength{\itemsep}{#2\baselineskip}}}
 {\end{list}}
\usepackage{calc}

\providecommand{\tightlist}{%
  \setlength{\itemsep}{0pt}\setlength{\parskip}{0pt}}

\usepackage{float}
\usepackage{tabularray}
\usepackage[normalem]{ulem}
\usepackage{graphicx}
\usepackage{rotating}
\UseTblrLibrary{booktabs}
\UseTblrLibrary{siunitx}
\NewTableCommand{\tinytableDefineColor}[3]{\definecolor{#1}{#2}{#3}}

\usepackage{float}
\KOMAoption{captions}{tableheading}
\makeatletter
\@ifpackageloaded{caption}{}{\usepackage{caption}}
\AtBeginDocument{%
\ifdefined\contentsname
  \renewcommand*\contentsname{Table of contents}
\else
  \newcommand\contentsname{Table of contents}
\fi
\ifdefined\listfigurename
  \renewcommand*\listfigurename{List of Figures}
\else
  \newcommand\listfigurename{List of Figures}
\fi
\ifdefined\listtablename
  \renewcommand*\listtablename{List of Tables}
\else
  \newcommand\listtablename{List of Tables}
\fi
\ifdefined\figurename
  \renewcommand*\figurename{Figure}
\else
  \newcommand\figurename{Figure}
\fi
\ifdefined\tablename
  \renewcommand*\tablename{Table}
\else
  \newcommand\tablename{Table}
\fi
}
\@ifpackageloaded{float}{}{\usepackage{float}}
\floatstyle{ruled}
\@ifundefined{c@chapter}{\newfloat{codelisting}{h}{lop}}{\newfloat{codelisting}{h}{lop}[chapter]}
\floatname{codelisting}{Listing}

\makeatother
\makeatletter
\@ifpackageloaded{caption}{}{\usepackage{caption}}
\@ifpackageloaded{subcaption}{}{\usepackage{subcaption}}
\makeatother
\usepackage{bookmark}
\IfFileExists{xurl.sty}{\usepackage{xurl}}{} % add URL line breaks if available
\hypersetup{
  pdftitle={Same Prompt, Different Outcomes: Evaluating the Reproducibility of Data Analysis by LLMs},
  pdfauthor={Jiaxin (Allyson) Cui; Rohan Alexander},
  colorlinks=true,
  linkcolor={blue},
  filecolor={Maroon},
  citecolor={Blue},
  urlcolor={Blue},
  pdfcreator={LaTeX via pandoc}}

\title{Same Prompt, Different Outcomes: Evaluating the Reproducibility
of Data Analysis by LLMs\thanks{We thank Monica Alexander and Jamie
Stafford for helpful suggestions. We gratefully acknowledge the
financial support of the Data Sciences Institute at the University of
Toronto and NSERC Alliance. Cui conducted this analysis while working at
the Investigative Journalism Foundation (IJF) and we gratefully
acknowledge the IJF for providing the appointments database. Contact:
\href{mailto:rohan.alexander@utoronto.ca}{\nolinkurl{rohan.alexander@utoronto.ca}}.}}
\author{Jiaxin (Allyson) Cui \and Rohan Alexander}
\date{February 15, 2026}
\begin{document}
\maketitle
\begin{abstract}
We systematically evaluate the reproducibility of data analysis
conducted by Large Language Models (LLMs). We evaluate two prompting
strategies, six models, and four temperature settings, with ten
independent executions per configuration, yielding 480 total attempts.
We assess the completion, concordance, validity, and consistency of each
attempt and find considerable variation in the analytical results even
for consistent configurations. This suggests, as with human data
analysis, the data analysis conducted by LLMs can vary, even given the
same task, data, and settings. Our results mean that if an LLM is being
used to conduct data analysis, then it should be run multiple times
independently and the distribution of results considered.
\end{abstract}

\newpage


\textbf{CRediT contributions:} \emph{Cui:} Conceptualization; Data
Curation; Formal Analysis; Investigation; Methodology; Software;
Visualization; Writing -- Original Draft Preparation; Writing -- Review
\& Editing. \emph{Alexander:} Conceptualization; Funding Acquisition;
Methodology; Project Administration; Supervision; Visualization; Writing
-- Original Draft Preparation; Writing -- Review \& Editing.

\textbf{AI usage:} AI was used to help with aspects of coding including
interacting with APIs, data cleaning, and creating figures and tables,
but all code and results were reviewed and evaluated by the authors. AI
was used to suggest related literature and review the drafted paper to
identify mistakes, inconsistencies, and awkward phrasing.

\textbf{Significance statement:} Large Language Models are increasingly
used to generate code for data analysis, but the consistency of their
estimates is less commonly examined. We run 48 different configurations
of model, prompt, and temperature, ten times each, on a realistic data
analysis task. We find that even identical configurations can produce
substantively different results. Even at temperature zero some estimates
lead to different conclusions about the same research question. These
findings show that a single LLM-generated data analysis is insufficient
and that repeated independent executions should be standard practice.

\newpage

\section{Introduction}\label{introduction}

One foundation of science is the ability to follow the same steps as
someone else and get the same outcome. This concept goes by different
names in different disciplines (Barba 2018) but one definition of
reproducibility is ``obtaining consistent results with same data and
methods'' (National Academies of Sciences, Engineering, and Medicine
2019). To improve computational reproducibility many journals now insist
code and data are shared (Stodden, Seiler, and Ma 2018; Vilhuber 2020).

Alongside those initiatives has been the exploration of what is called
the ``many-analysts'' approach; where different groups of researchers
independently analyze the same dataset and compare their findings. For
instance, 29 teams analyzed the same data about whether soccer referees
give dark-skin-toned players red cards at higher rates, and found 9 of
the teams did not find a statistically significant effect, while the
other 20 did (Silberzahn et al. 2018). Other studies have found similar
differences between research teams (Botvinik-Nezer et al. 2020;
Ostropolets et al. 2023; Ortloff et al. 2023; Huntington-Klein et al.
2025). One key concern is that many of the differences come from
``\ldots differences in data preparation and analysis decisions, many of
which would not likely be reported in a publication'' (Huntington-Klein
et al. 2021, 944).

Large Language Models (LLMs) can now write code that outperforms or is
indistinguishable from, at least in some dimensions, that of humans (Hou
and Ji 2024; Xu et al. 2025). And there were early signs that LLMs could
use this capacity to write code to perform some aspects of data analysis
(Bubeck et al. 2023) which is now being explored in more realistic
settings (Li et al. 2026). As LLMs increasingly acquire the ability to
work with agency, that is without direct instruction to pursue goals
(Grootendorst and Alammar 2026), it will be the case that they
autonomously carry out data preparation and analysis.

In this paper we are interested in whether, like teams of humans, when
given the same task and data, LLMs come to different conclusions when
run multiple times. To date, many LLM evaluations consider best
performance rather than the distribution of performance. This is
appropriate in fields such as software engineering, where the focus may
be on code passing appropriate tests regardless of how many attempts are
required. For instance, the \texttt{pass@k} metric (Kulal et al. 2019)
was used to measure whether at least one of \(k\) functions generated
from a docstring by the language model Codex was correct (Chen et al.
2021). LLMs are foundationally statistical, and can exhibit
non-determinism even when temperature is zero (Atil et al. 2025; He and
Lab 2025). In practice this means there is almost always some chance
that running the same prompt repeatedly will result in a different
outcome. This is one reason that the \texttt{pass@k} metric is so useful
in those fields; achieving a certain level of performance once, means it
has been achieved, even if other attempts fall short.

LLMs are now used to generate analysis code that underpins scientific
claims in a wide-variety of areas including infectious disease (Kraemer
et al. 2025), scientific discovery (Wang et al. 2023), and materials
discovery (Cheng et al. 2026). In contrast to code in a software
engineering environment, it may be that code used for scientific
discovery and data analysis is not run too many times if it is just used
as the basis of a finding. But if variation is important, as it is when
we are interested in whether some difference is statistically
significant, then it is not necessarily sufficient that the data
analysis conducted by an LLM results in a particular conclusion once.

We examine whether the same prompt, executed multiple times, results in
substantially different conclusions in a realistic data analysis task.
We evaluate six models from three providers: Claude Haiku 4.5 and Claude
Opus 4.6 from Anthropic, GPT-5-mini and GPT-5.2 from OpenAI, and Gemini
Flash 3 and Gemini Pro 3 from Google. We consider four temperature
settings: 0.0, 0.3, 0.7, and 1.0, with lower settings being associated
with more highly probable tokens being returned (Alammar and
Grootendorst 2024, 171). And we consider two different prompt
approaches: single-step and multi-step. In the former, we use one prompt
to produce a complete analysis pipeline. In the latter we use a series
of sequential prompts to complete individual stages with intermediate
validation. Multi-step prompts should allow course correction, while
single-step prompts are more closely related to how human data analysts
work in the real world and speak to potential economic benefits from AI.
For each model and setting we consider ten independent executions for
each prompting strategy. This results in 480 total execution attempts.

We evaluate performance on a five-step data analysis pipeline processing
government appointment records for a Canadian province: consolidate
multiple data files, identify reappointment patterns, aggregate to
analytical units, perform OLS regression, and generate visualizations.
The task is similar to that done by entry-level data scientists or
junior analysts.

We document three main findings. First, single-step prompting tends to
have higher completion rates than multi-step prompting, because in the
multi-step approach an error at any stage prevents all later stages from
running. Second, the generated code is almost always structurally valid
but can differ from the human-conducted analysis, especially in the data
preparation steps, where choices about sorting, missing values, and
which records to include are underspecified. Third, these data
preparation differences propagate to the analysis, producing distinct
clusters of estimates. While most configurations produce t-statistics
consistent with no statistically significant relationship, the estimated
slopes vary in both sign and magnitude. This means the same data and
task can lead to different conclusions depending on which analytical
pathway the model follows.

\section{Results}\label{results}

\subsection{Overall}\label{overall}

\begin{figure}

\centering{

\pandocbounded{\includegraphics[keepaspectratio]{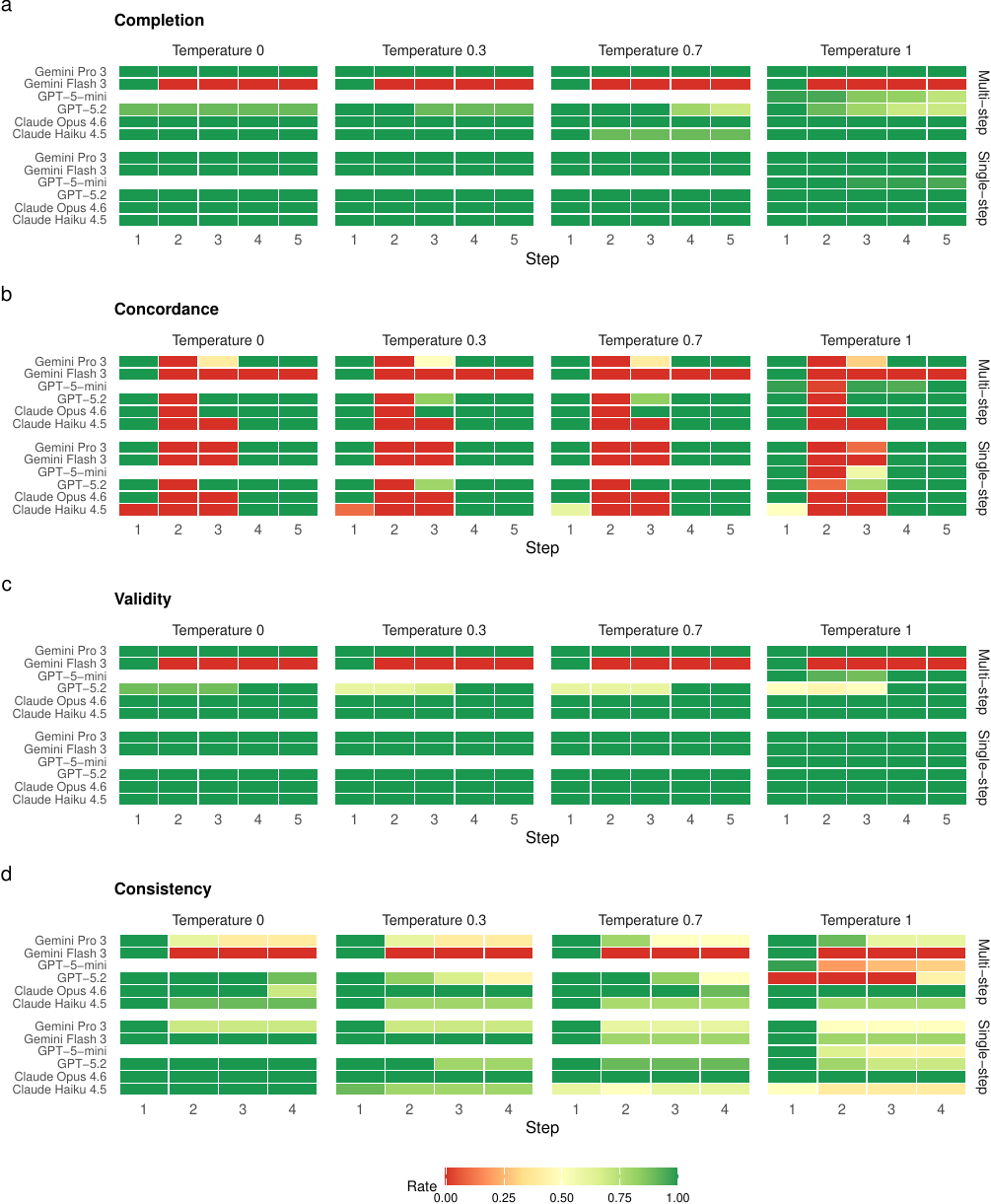}}

}

\caption{\label{fig-eval-heatmap}Evaluation metrics across pipeline
steps, models, temperatures, and prompting strategies. Each tile shows
the rate for one model-step combination. Rows are grouped by prompting
strategy, columns by temperature. Color intensity indicates the metric
value from 0 (red) to 1 (green). GPT-5-mini does not support custom
temperature and is only evaluated at its default temperature of 1.}

\end{figure}

We evaluate the LLM execution attempts on four criteria: completion,
concordance, validity, and consistency (Figure~\ref{fig-eval-heatmap}).
Completion is whether the generated code executes without errors and
produces the expected output files. Concordance compares the output to a
human-conducted analysis. Validity is whether the outputs have the
expected columns, names, data types, and range of values. And
consistency is the similarity between each of the 10 executions of a
particular configuration. Figure~\ref{fig-eval-strat} in
Appendix~\ref{sec-additionalresults} provides additional detail.

\subsection{Completion}\label{completion}

Figure~\ref{fig-eval-heatmap} (a) shows completion rates across pipeline
steps, temperatures, and models for both prompting strategies.
Completion measures whether the generated code produces all expected
output files.

Figure~\ref{fig-eval-heatmap} (a) shows that the single-step prompt
approach, where the model receives all the instructions at once, tends
to have higher completion rates than the multi-step prompt approach,
where tasks are broken into smaller prompts. This may seem
counterintuitive. One may have thought that breaking up a task into
discrete steps would make it easier for the model to achieve. But
Figure~\ref{fig-eval-heatmap} (a) shows the opposite because of error
propagation.

In the multi-step prompt approach, each step produces a separate script
that must execute successfully before the next step can run. An error at
any point propagates; if, say, Step 2 fails, then Steps 3 to 5 do not
execute even if they would have worked. In the single-step prompt
approach the model is able to produce one internally consistent script.
There is less opportunity for propagating failures between steps because
all code is generated and executed as a single unit. GPT-5-mini and
Gemini Flash 3 illustrate this, with considerable differences between
the multi-step and single-step completion rates. Gemini Flash 3 achieves
100 per cent completion in single-step mode but 0 per cent in multi-step
mode because every execution fails at Step 2 due to incorrect file path
handling.

Completion rates are high for most models across all temperature values
for the single-step prompt. Temperature does not appear to have a
consistent relationship with completion rates for the multi-step prompt
either, though GPT-5-mini does not support custom temperature settings.

\subsection{Concordance}\label{concordance}

There are a variety of ways to consider concordance between the LLM
estimates and those of a human. One way is to look at concordance across
all steps. Concordance rates in Figure~\ref{fig-eval-heatmap} (b) are
based on executions that completed successfully. A model that failed to
produce output files for a given step is excluded from the concordance
calculation for that step. For instance, some configurations of
GPT-5-mini with multi-step prompting had only 60 per cent completion at
Step 5, even though the executions that did complete Step 5 can have
high concordance rates. This pattern suggests that the model can produce
concordant output when it works, but frequently fails to produce working
code at all.

\begin{figure}

\centering{

\pandocbounded{\includegraphics[keepaspectratio]{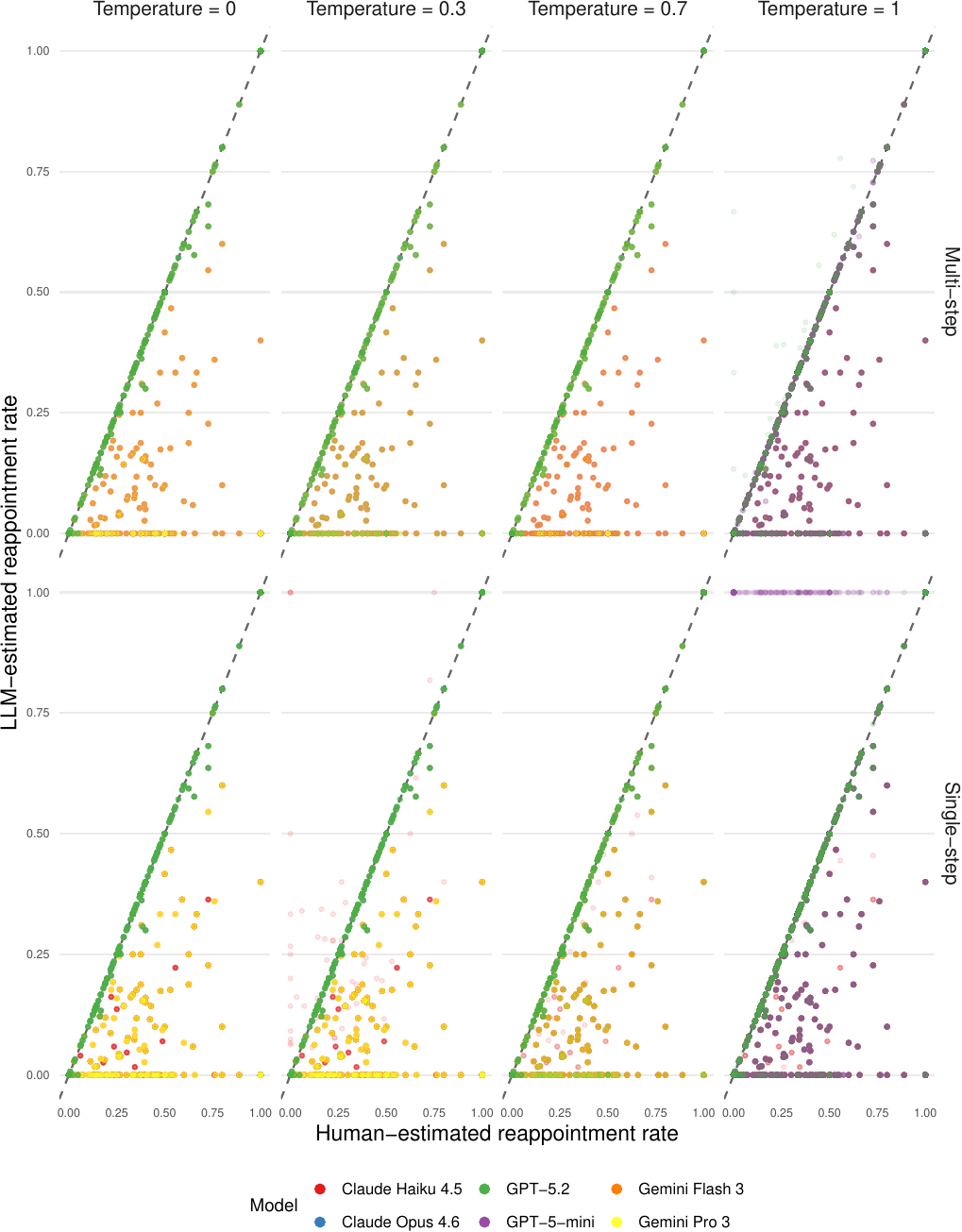}}

}

\caption{\label{fig-reappointment-scatter}Comparison of LLM-estimated
reappointment rates to those from human analysis at the department-year
level. Each point is one department-year observation from one execution.
The dashed 45-degree line indicates the estimates are the same.
GPT-5-mini is only evaluated at its default temperature of 1.}

\end{figure}

Another way to consider concordance is to focus on only one aspect. For
instance, take the estimated annual reappointment rate for each
department. If we compare the rates generated by the LLMs with those
generated by a human we find considerable differences
(Figure~\ref{fig-reappointment-scatter}). With only a few exceptions,
almost all the differences between the LLM estimate of the
re-appointment rate and the actual re-appointment rate, are lower.

Figure~\ref{fig-reappointment-scatter} compares individual reappointment
rate estimates at the organization-year level and shows that many LLM
outputs are the same or close to the actual rate.
Figure~\ref{fig-eval-heatmap} (b) is stricter---a value is concordant
only if it is identical (after rounding)---any difference in values,
structure, or which observations are included, is considered
non-concordant. The concordance rate in Figure~\ref{fig-eval-heatmap}
(b) is the proportion of the ten executions per configuration that
achieve an exact match. And so the difference between
Figure~\ref{fig-reappointment-scatter} and Figure~\ref{fig-eval-heatmap}
(b) is because of that different level of granularity.
Figure~\ref{fig-eval-heatmap} (b) shows where the models do better and
worse. Steps 2 and 3 consistently have the lowest concordance rates, and
Steps 4 and 5 tend to have higher concordance rates. It might be that
regression computation and visualization are well-specified tasks and so
the model is able to match the human outputs more easily. Step 5
concordance only checks whether a PNG file was produced, not whether the
image content matches, so its high rates should be interpreted as
completion rather than visual similarity.

\subsection{Validity}\label{validity}

Figure~\ref{fig-eval-heatmap} (c) shows validity rates conditional on
file existence. Validity assesses whether produced outputs satisfy task
specifications: correct column presence and naming, appropriate data
types, values within valid ranges. The validity rate is high for almost
all configurations.

\subsection{Consistency}\label{consistency}

Figure~\ref{fig-eval-heatmap} (d) shows consistency across repeated
executions within each configuration. For each configuration we identify
the most common output across the ten executions and calculate the
fraction that match it. A rate of 1.0 means all executions that produced
output for that step were identical; lower values indicate that
executions diverged. Step 5 consistency is not assessed because
visualization outputs are not numerically compared. The results show
that even when temperature is zero it cannot be assumed that the model
will provide the same results.

\begin{figure}

\centering{

\pandocbounded{\includegraphics[keepaspectratio]{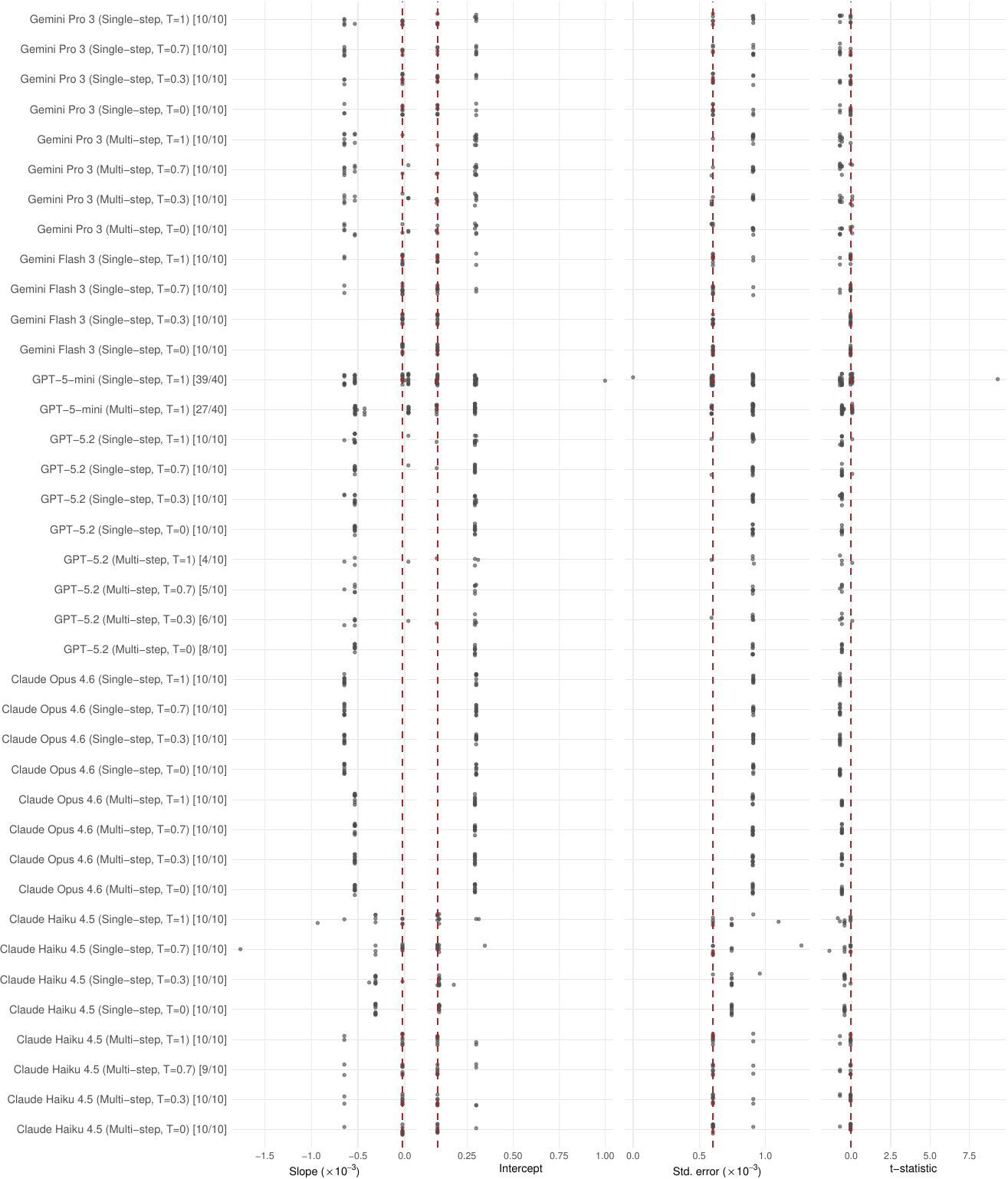}}

}

\caption{\label{fig-execution-dotplot}Output from each individual
execution, by configuration. Each dot is one independent execution. The
number in brackets shows how many executions produced valid regression
output out of the total attempts. GPT-5-mini configurations are
collapsed across temperature settings because the model does not support
custom temperature. Configurations are grouped by model, then prompt
strategy, then temperature. The dashed vertical line shows the
human-generated value.}

\end{figure}%

Figure~\ref{fig-execution-dotplot} shows the regression slope,
intercept, standard error, and t-statistic from each individual
execution that completed Step 4. Each dot represents one execution and
configurations are grouped by model, then prompt strategy, then
temperature, with that ordering held constant across all four panels so
that the same row refers to the same configuration throughout. The
number in brackets on the y-axis shows how many executions produced
valid regression output. Because GPT-5-mini does not support custom
temperature, its four temperature groups are collapsed into a single row
per strategy with 40 total attempts. Configurations with fewer completed
executions, such as GPT-5-mini (Multi-step, T=1) with 33 out of 40,
appear sparse not because of consistency but because some attempts
failed before producing a regression. The dashed vertical line indicates
the human-generated value.

That there are particular clusters of estimates may suggest the LLMs
follow particular analytical pathways and that data preparation choices
are having a large effect on the outcome. The standard error panel shows
differences between configurations about precision. This may suggest
differences in the number of observations retained after data cleaning.
Most executions produce t-statistics near zero, consistent with the
human-analysis, and none are above two in absolute value for multi-step
prompting, though a small number of single-step configurations produce
t-statistics that could lead to a conclusion of statistically
significant differences.

\section{Discussion}\label{discussion}

We evaluate LLM code generation reliability and consistency across two
prompting strategies, six models, and four temperature settings. We find
that output differs even between executions with the same settings, and
that the results can be different to that of a human. That said, the
LLMs almost always produce structurally valid code, especially when the
overall task was presented to them at once, rather than in pieces. Code
and data sharing was expected to allow the validation of methodology
across workflows (Donoho 2017) and we show that recent advances in AI
have moved us closer to that future.

Our results demonstrate the need to run many executions of the same task
when conducting data analysis with LLMs. Different models performed
differently, and so our results also emphasize the importance of running
any data analysis with models from at least two different providers. At
the very least it is important to show any variability in the results if
the focus is inference, but it also might be possible to consider some
type of ensemble approach if the focus is prediction.

That all said, the estimates from the LLMs tended to differ in
particular ways. This suggests the LLMs generated code that resulted in
differences at particular points in the analysis workflow. This is
similar to findings in the context of many-analyst studies who conclude
that many of the differences are due to different choices in terms of
data preparation and cleaning (Huntington-Klein et al. 2025). Like
earlier work in the context of human many-analyst analysis (Silberzahn
et al. 2018), these structured differences result in different estimated
slope coefficients.

The differences tended to occur at specific pipeline stages. Some models
exclude files, or parse years differently. Temporal ordering and
tie-breaking produce variation at Step 2, with different stable sorts
changing which records count as first occurrences. Missingness handling
determines whether null organization values persist through aggregation,
directly affecting regression sample composition. This suggests a
possible lack of specificity in the prompts. But that is the reality of
data analysis, which is an interactive process of learning from data. We
show that two AI agents could autonomously reach different conclusions
about the same task from the same data. As the cost of agentic data
analysis decreases, this suggests a need, as discussed in the context of
human many-analyst studies (Auspurg and Brüderl 2024), for guidelines to
enable systematic comparison as it becomes cheaper to explore the
``hidden universe of uncertainty'' (Breznau et al. 2022).

Our study has several limitations. We evaluate a single analytical task
in one domain using one dataset, and the extent to which our findings
generalize to other tasks, disciplines, or data structures is unknown.
Our concordance measure relies on a human analyst, but the many-analysts
approach suggests multiple valid approaches may exist for the same task.
That said, we are not necessarily saying results that differ to those
generated by a human are wrong, just that they are different. We test
six models from three providers at one point in time, but model behavior
changes with updates, and our results are a snapshot. We consider only
single-shot code generation without iteration, but in practice analysts
review and revise LLM outputs, and agentic frameworks that allow models
to execute, observe errors, and retry may produce different results.

It would be interesting to consider different data structures and
domains, as well as whether more or less specific prompts would affect
our results. It would be especially interesting to consider agentic
workflows, where the model can inspect intermediate outputs and correct
errors before proceeding. Most importantly, our work points to the need
to develop methods to automatically enable multiple, independent,
executions and identify they have converged on different analytical
pathways, and to consider the distribution of estimates.

\section{Method}\label{method}

The underlying substantive dataset is annual public appointment records
in a Canadian province, New Brunswick. These are real data from the
Investigative Journalism Foundation. They contain the normal messiness
found in real-world data, such as inconsistent names, missing values,
and errors. Additional details about the data are provided in
Appendix~\ref{sec-underlying-data}. We want to answer the research
question: How does the rate of reappointment in New Brunswick government
departments differ between departments and change over time? To do this
we establish a five-step data processing and analysis pipeline:

\begin{enumerate}
\def\labelenumi{\arabic{enumi}.}
\tightlist
\item
  Bring together the 12 annual (2013 to 2024, inclusive) CSV files into
  one dataset.
\item
  Identify individual reappointments by matching
  name-position-organization combinations.
\item
  Aggregate individual reappointments to organization-year summaries
  computing total appointments, reappointments, and rates.
\item
  Estimate a regression with the reappointment rate as the outcome
  variable and total appointments as the predictor variable.
\item
  Create a scatter plot with a regression line overlay.
\end{enumerate}

Each step is described in natural language similar to how a manager may
instruct a junior data analyst. This means that not all details are
fully specified and some variation is possible.

We consider two approaches to prompting. In the first approach,
``multi-step'', models sequentially receive one prompt for each of the
five steps. Each prompt specifies the step's task, expected input files,
and required output. The LLM generates individual Python scripts that
are then run sequentially. In the second approach, ``single-step'', the
LLM gets one prompt that contains all five steps and their inputs and
outputs. The LLM is asked for one Python script that will execute all
aspects. Both prompting approaches have the same task specifications
and, if successful, have the same outputs. Additional details about our
prompts are provided in Appendix~\ref{sec-prompts}.

We evaluated six models from three providers: Claude Haiku 4.5 and
Claude Opus 4.6 (Anthropic), GPT-5-mini and GPT-5.2 (OpenAI), and Gemini
Flash 3 and Gemini Pro 3 (Google). Each model was tested with two prompt
strategies, at four temperature settings, 0.0, 0.3, 0.7, 1.0, with ten
independent executions for each configuration. This resulted in 480
total execution attempts. GPT-5-mini does not support custom temperature
settings, so its executions use the model's default temperature of 1 for
40 executions (OpenAI 2026). GPT-5.2 requires disabling built-in
reasoning to accept a temperature parameter. Gemini Pro 3 requires a
thinking mode that cannot be disabled; we set a minimal thinking budget
to approximate standard generation. By way of background, temperature is
associated with how likely it is that less probable tokens are returned,
and a temperature of zero means highly probable tokens are returned,
while a temperature of one means sometimes less probable tokens are
returned (Alammar and Grootendorst 2024, 171).

We conducted all analysis and produced all figures and tables using the
statistical programming language \texttt{R} (R Core Team 2025), with the
\texttt{tidyverse} (Wickham et al. 2019), and \texttt{tinytable}
(Arel-Bundock 2025) packages.

\newpage

\appendix

\section*{Appendix}\label{appendix}
\addcontentsline{toc}{section}{Appendix}

\setcounter{figure}{0}
\renewcommand{\thefigure}{A\arabic{figure}}

\setcounter{table}{0}
\renewcommand{\thetable}{A\arabic{table}}

\section{Underlying data}\label{sec-underlying-data}

Our underlying data are released through Orders-in-Council (OiCs) in
Canadian provinces. OiCs formalize appointments to government agencies,
boards, and commissions, creating administrative records of public
staffing decisions. The Investigative Journalism Foundation compiles
these records into standardized ``appointments databases'' documenting
appointment attributes including appointee name, organization, position,
appointment dates, and reappointment status (Investigative Journalism
Foundation 2025). We focus on New Brunswick province.
Table~\ref{tbl-data-summary} presents summary statistics.

\begin{table}[H]

\caption{\label{tbl-data-summary}Summary statistics for the New
Brunswick appointments dataset (2013--2024). The dataset contains 3,718
appointment records across 43 organizations. Reappointment rates are
calculated as the proportion of records flagged as reappointments in the
original administrative data.}

\centering{

\centering
\begin{tblr}[         %% tabularray outer open
]                     %% tabularray outer close
{                     %% tabularray inner open
colspec={Q[]Q[]},
cell{1}{1}={}{halign=l, font=\bfseries,},
cell{1}{2}={}{halign=r, font=\bfseries,},
cell{2-12}{1}={}{halign=l,},
cell{2-12}{2}={}{halign=r,},
}                     %% tabularray inner close
\toprule
Characteristic & Value \\ \midrule %% TinyTableHeader
Total appointments (2013-2024) & 3,718 \\
Unique appointees & 2,357 \\
Unique organizations & 43 \\
Years covered & 12 \\
Appointments per year (mean) & 310 \\
Appointments per year (SD) & 72 \\
Organizations per year (mean) & 17 \\
Organizations per year (SD) & 3 \\
Records with missing organization & 1.5\% \\
Records marked as reappointments & 874 \\
Reappointment rate (aggregate) & 23.5\% \\
\bottomrule
\end{tblr}

}

\end{table}%

Table~\ref{tbl-raw-example} is an extract from the raw appointment
records. This shows the structure of the data that the models must
analyze.

\begin{table}[H]

\caption{\label{tbl-raw-example}Example records from the New Brunswick
appointments database (2013). Fields include appointee name, position,
organization, reappointment status, and administrative metadata. Records
show variation in completeness and formatting typical of administrative
data.}

\centering{

\centering
\begin{tblr}[         %% tabularray outer open
]                     %% tabularray outer close
{                     %% tabularray inner open
width={1\linewidth},
colspec={X[]X[]X[]X[]X[]},
row{1}={}{halign=l, font=\bfseries, font=\fontsize{0.8em}{1.1em}\selectfont,},
row{2,3,4,5,6,7}={}{halign=l, font=\fontsize{0.8em}{1.1em}\selectfont,},
}                     %% tabularray inner close
\toprule
Name & Position & Org & Reappointed & Start Date \\ \midrule %% TinyTableHeader
Cindy Howe & members & Health & FALSE & 2013-01-09 \\
Floyd Haley & members & Health & FALSE & 2013-01-09 \\
Frank B. Trevors & members & Health & FALSE & 2013-01-09 \\
Gerald Richard & Deputy Minis... & Executive Co... & FALSE & 2013-01-21 \\
Harry Doyle & members & Health & FALSE & 2013-01-09 \\
Jeffrey Beairsto & members & Health & FALSE & 2013-01-09 \\
\bottomrule
\end{tblr}

}

\end{table}%

\begin{figure}

\centering{

\pandocbounded{\includegraphics[keepaspectratio]{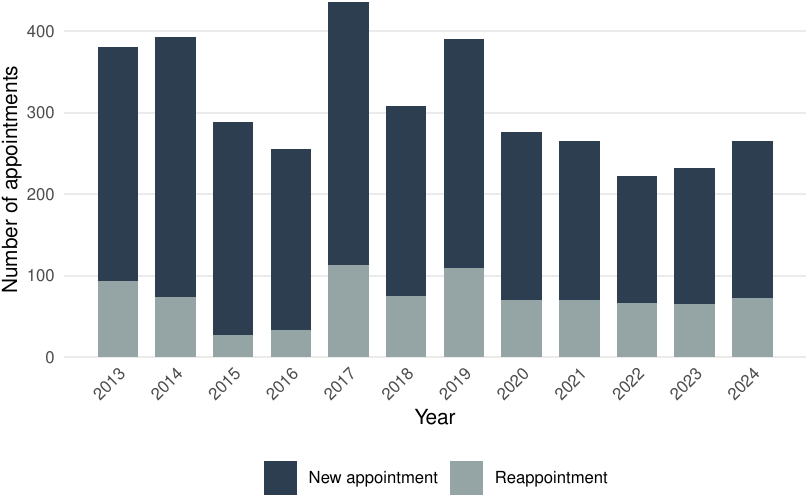}}

}

\caption{\label{fig-appointments}Annual appointment counts by
reappointment status in New Brunswick (2013--2024). Dark bars show new
appointments while light bars show reappointments. There is year-to-year
variation in both total appointments and the number of reappointments.}

\end{figure}

Figure~\ref{fig-appointments} shows the distribution of appointments
over time. These are the numbers that we use to compare to the estimates
from the models.

\newpage

\section{Prompt details}\label{sec-prompts}

\subsection{Multi-step prompts}\label{multi-step-prompts}

The multi-step prompt approach uses five sequential prompts. Each prompt
is preceded by:

\begin{verbatim}
Only respond with the python script that can be directly used to run the task,
do not include any explanation or other text. Therefore, your response should
start with "import...", and end with "if __name__ == '__main__': main()"
\end{verbatim}

\textbf{Step 1:} The prompt specifies multiple CSV files located at
\texttt{prompt\_raw\_data/appointments\_\{year\}.csv}, each containing
16 columns (name, position, org, location, region, posted\_date,
start\_date, end\_date, term\_length, acts, remuneration, reappointed,
oic, href, body, link). The model is asked to combine the files, add a
year column, and keep only name, position, org, reappointed, and year.
The expected output is \texttt{step1.csv} with those five columns, all
nullable.

\textbf{Step 2:} The prompt provides \texttt{step1.csv} as input and
asks the model to sort by name, position, org, year, and reappointed,
then update the reappointed column to true if the same name-position-org
combination appeared for not the first time. The expected output is
\texttt{step2.csv} with the same five columns.

\textbf{Step 3:} The prompt provides \texttt{step2.csv} as input and
asks the model to remove the name and position columns, compress the
dataset to year-org groups with total appointments and total
reappointments, and add a reappointment rate column. The expected output
is \texttt{step3.csv} with five columns: year, org, total\_appointments,
total\_reappointments, and reappointment\_rate.

\textbf{Step 4:} The prompt provides \texttt{step3.csv} as input and
asks the model to perform an OLS regression of reappointment\_rate on
total\_appointments, document the slope, intercept, standard error, and
t-statistics, and comment on statistical and economic significance. The
expected output is \texttt{step4.csv} with eight columns including the
regression coefficients, significance flags, and text comments.

\textbf{Step 5:} The prompt provides both \texttt{step3.csv} and
\texttt{step4.csv} as input and asks the model to create a scatter plot
with total\_appointments on the x-axis and reappointment\_rate on the
y-axis, draw a regression line using the slope and intercept, and use a
viridis colormap by year. The expected output is \texttt{step5.png}.

\subsection{Single-step prompt}\label{single-step-prompt}

The single-step prompt approach just brings together all five steps into
one prompt with the same prefix. So the task descriptions are combined
and it is specified that a single complete pipeline script should be
generated.

\section{Additional result details}\label{sec-additionalresults}

\begin{figure}

\begin{minipage}{0.50\linewidth}

\centering{

\pandocbounded{\includegraphics[keepaspectratio]{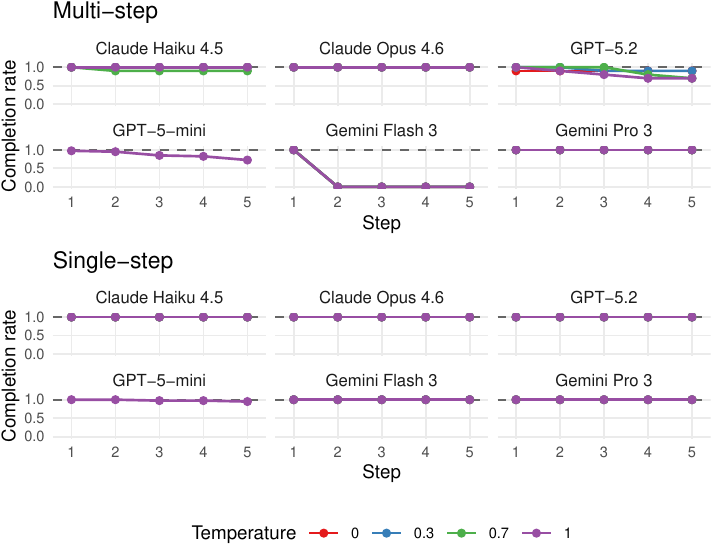}}

}

\subcaption{\label{fig-eval-strat-1}Completion}

\end{minipage}%
\begin{minipage}{0.50\linewidth}

\centering{

\pandocbounded{\includegraphics[keepaspectratio]{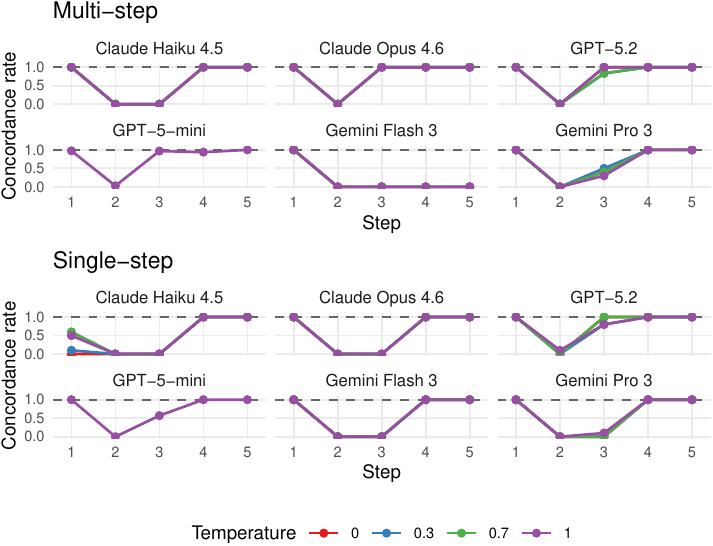}}

}

\subcaption{\label{fig-eval-strat-2}Concordance}

\end{minipage}%
\newline
\begin{minipage}{0.50\linewidth}

\centering{

\pandocbounded{\includegraphics[keepaspectratio]{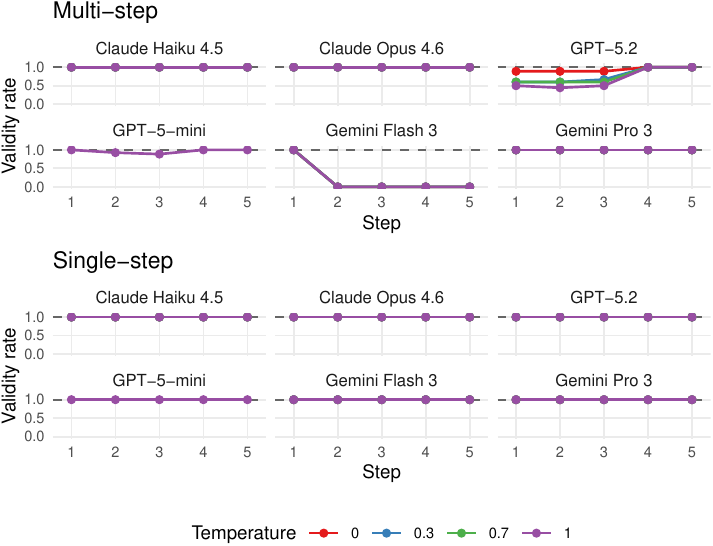}}

}

\subcaption{\label{fig-eval-strat-3}Validity}

\end{minipage}%
\begin{minipage}{0.50\linewidth}

\centering{

\pandocbounded{\includegraphics[keepaspectratio]{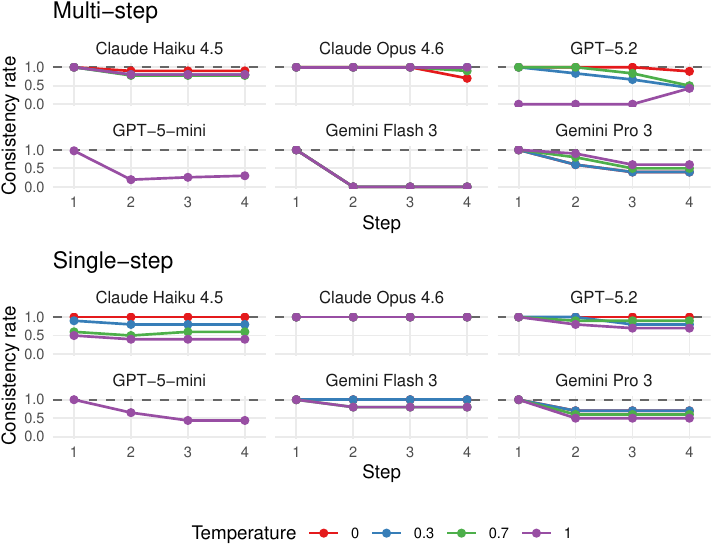}}

}

\subcaption{\label{fig-eval-strat-4}Consistency}

\end{minipage}%

\caption{\label{fig-eval-strat}Evaluation metrics across pipeline steps,
faceted by model. Each panel shows one model's trajectory across all
temperatures. Line color indicates temperature. Top two rows show
multi-step, bottom two show single-step. GPT-5-mini is only evaluated at
its default temperature of 1.}

\end{figure}%

\newpage

\section*{References}\label{references}
\addcontentsline{toc}{section}{References}

\phantomsection\label{refs}

\end{document}